\documentclass[aps,prl,twocolumn,showpacs,superscriptaddress,groupedaddress]{revtex4}

\usepackage{graphicx}  
\usepackage{dcolumn}   
\usepackage{bm}        
\usepackage{amsmath} 
\usepackage{amssymb}   
\usepackage[caption=false]{subfig}
\usepackage{hyperref}
\usepackage{comment}
\usepackage{color}
\usepackage{url}

\usepackage[normalem]{ulem}

\begin{document}
\title{Electron acceleration by relativistic surface plasmons in laser-grating interaction}

\author{L. Fedeli}
\email{luca.fedeli@for.unipi.it}
\affiliation{Enrico Fermi Department of Physics, University of Pisa, Pisa, Italy}
\affiliation{National Institute of Optics, National Research Council (CNR/INO), u.o.s Adriano Gozzini, Pisa, Italy}

\author{A. Sgattoni}
\affiliation{National Institute of Optics, National Research Council (CNR/INO), u.o.s Adriano Gozzini, Pisa, Italy}

\author{G. Cantono}
\affiliation{CEA/DSM/IRAMIS/LIDYL, Gif-sur-Yvette, France}
\affiliation{University of Paris Sud, Orsay, France}
\affiliation{Enrico Fermi Department of Physics, University of Pisa, Pisa, Italy}
\affiliation{National Institute of Optics, National Research Council (CNR/INO), u.o.s Adriano Gozzini, Pisa, Italy}

\author{D. Garzella}
\author{F. R\'eau}
\affiliation{CEA/DSM/IRAMIS/LIDYL, Gif-sur-Yvette, France}

\author{I. Prencipe}
\author{M. Passoni}
\affiliation{Department of Energy, Politecnico di Milano University, Milan, Italy}

\author{M. Raynaud}
\affiliation{Laboratoire des Solides irradi\'es, Ecole Polytechnique, CNRS, CEA/DSM/IRAMIS, Universit\'e Paris-Saclay, Palaiseau Cedex, France}

\author{M. Kv\v{e}to\v{n}}
\author{J. Proska}
\affiliation{FNSPE, Czech Technical University, Prague, Czech Republic}

\author{A. Macchi}
\affiliation{National Institute of Optics, National Research Council (CNR/INO), u.o.s Adriano Gozzini, Pisa, Italy}
\affiliation{Enrico Fermi Department of Physics, University of Pisa, Pisa, Italy}

\author{T. Ceccotti}
\affiliation{CEA/DSM/IRAMIS/LIDYL, Gif-sur-Yvette, France}

\begin{abstract}
The generation of energetic electron bunches by the interaction of a short, ultra-intense (${I~>10^{19} \:\textrm{W/cm}^2}$) laser pulse with ``grating'' targets has been investigated in a regime of ultra-high pulse-to-prepulse contrast (∼$10^{12}$).
For incidence angles close to the resonant condition for Surface Plasmon (SP) excitation, a strong electron emission was observed within a narrow cone along the target surface, with energies exceeding 10~MeV. Both the energy and the number of emitted electrons were strongly enhanced with respect to simple flat targets.
The experimental data are  closely reproduced by three-dimensional particle-in-cell simulations, which provide evidence for the generation of relativistic SPs and for their role in driving the acceleration process. Besides the possible applications of the scheme as a compact, ultra-short source of MeV electrons, these results are a step forward the development of high field plasmonics.
\end{abstract}


\maketitle

Surface plasmons (SP) are electromagnetic modes which can be excited at a sharp material interface having a periodic modulation, e.g. a grating, to allow phase matching with incident laser light. SP excitation allows enhancement and fine control of electromagnetic field coupling with structured metal surfaces, which is the basis for the vibrant research field of plasmonics \cite{plasmonicsEdiorial} with several applications ranging from light concentration beyond diffraction limit \cite{plasmonics2} to biosensors \cite{plasmonics3} and plasmonic chips \cite{Ozbay13012006}.

Extending the study of plasmonics to very high fields might allow novel applications, such as tapered wave-guides for high energy concentration and field amplification to extreme values \cite{Esmann2013,PhysRevLett.93.137404} or enhanced energy absorption which could be exploited, e.g., for proton acceleration by integrating the gratings in complex target assemblies (see e.g. \cite{:/content/aip/journal/pop/18/5/10.1063/1.3575624,bartalNP12}). In particular, the phase velocity and the longitudinal field component of SPs make them suitable for accelerating electrons to high energies, provided that the SP field is driven high enough for the electron dynamics to become relativistic. 
While a theory of relativistic SP is still lacking, numerical simulations have provided evidence of SP-related effects in the relativistic regime 
\cite{macchiPRL01,bigongiariPoP13} 
including electron acceleration in grating targets at weakly relativistic intensities
\cite{:/content/aip/journal/pop/14/9/10.1063/1.2755969}.
However,  most experiments so far have been restricted to intensities below $10^{16} ~  \textrm{W/cm}^2$ \cite{:/content/aip/journal/pop/17/3/10.1063/1.3368792, :/content/aip/journal/pop/17/8/10.1063/1.3469576, :/content/aip/journal/pop/19/3/10.1063/1.3693388}, unable to drive a relativistic SP, because of the intense prepulses inherent to high-power laser systems which can lead to an early disruption of the target structuring. 
The development of devices for ultra-high contrast pulses \cite{ucont1,ucont2} now allows to explore the interaction with solid targets structured on a sub-micrometric scale in the relativistic regime \cite{PhysRevLett.110.065003,purvisNatPhot13}. In particular, an enhancement of the cut-off energy of protons accelerated from the rear surface of grating targets was observed and related to SP excitation \cite{PhysRevLett.111.185001}.

In this Letter, we demonstrate electron acceleration with relativistic SPs excited on the surface of grating targets at laser intensities $I=5 \times 10^{19} ~ \textrm{W/cm}^2$, corresponding to a relativistic parameter $a_0 \simeq 5$ where $a_0=(I\lambda^2/10^{18}~\textrm{W cm}^{-2}\mu\textrm{m}^2)^{1/2}$. With respect to flat targets, we observed a strong enhancement of both the energy and number of electrons emitted from gratings irradiated at an incidence angle close to the resonant value for SP excitation. Electron emission was concentrated in a narrow cone close to the target surface, with energies  exceeding 10~MeV.

The basics of SP generation and electron acceleration may be described as follows. At high laser intensities ($I>10^{18} ~ \textrm{W/cm}^2$) a solid target is ionized within one laser cycle, thus the interaction occurs with a dense plasma. Assuming a plasma dielectric function $\varepsilon(\omega)=1-\omega_p^2/\omega^2 \equiv 1-\alpha$ (where $\omega_p$ is the plasma frequency) the phase velocity of a SP is $v_f=\omega/k=c(\alpha-2)^{1/2}/(\alpha-1)^{1/2}$ where $k$ is the SP wave vector and $\alpha>2$ holds.  The condition for resonant excitation of a SP on a periodically modulated target (grating) by an incident EM wave 
of the same frequency is $\lambda/\lambda_g = {(1 - \alpha)^{1/2}/(2-\alpha)^{1/2}}-\sin(\phi_i)$ where $\lambda_g$ is the grating period and $\phi_i$ is the angle of incidence. Notice that these equations neglect thermal and collisional effects as well as possible relativistic non-linearities, thus in principle the resonance could be expected at somewhat different angles. 

Electron acceleration  up to relativistic energy by the longitudinal field of a SP requires the phase velocity $v_f$ to approach $c$ in order to minimize dephasing, thus $\alpha\gg 1$ is required as expected for a solid-density plasma and optical frequencies. The basic process may be described very similarly to the well-known acceleration in wake plasma waves \cite{tajima}, but with the difference that the transverse field component of the SP pushes electron on the vacuum side, so that the process is two-dimensional and eventually the electrons are emitted at some angle with respect to the SP propagation direction.
In a frame $L'$ moving with velocity ${\bf v}_f=v_f\hat{\bf y}$
with respect to the laboratory  frame $L$, the SP field is electrostatic in the vacuum region ($x<0$) and can be derived from the potential
\begin{eqnarray}
\Phi'=- \frac{{E_{\mbox{\tiny SP}}}}{{k'}} \mbox{e}^{k'x}\sin{k'y'} \; ,  
\label{eq:theory1}
\end{eqnarray}
where $\gamma_f=(1-v_f^2/c^2)^{-1/2}=(\alpha-1)^{1/2}$, $k'=k/\gamma$, and $E_{\mbox{\tiny SP}}$ is the amplitude of the longitudinal SP field ($E_y$) in $L$. 
Thus, in the $L'$ frame the process is simply described as the electron going downhill the potential energy $-e\Phi'$. 
Due to the evanescent field component $E'_x=-\partial_x\Phi'$, electrons are predominantly accelerated towards the vacuum side with velocity almost normal to the $x=0$ surface. The condition of ``optimal'' injection corresponds to an electron placed initially in $L'$ at the top of the potential hill ($x'=0,\, y'=\pi/2k'$) with $v'_y=0$, i.e. with an initial velocity ${\bf v}_f$ in $L$. Such electron will acquire in $L'$ the energy $W'=eE_{\mbox{\tiny SP}}\gamma_f/k$. If $a_{\mbox{\tiny SP}}\equiv eE_{\mbox{\tiny SP}}/m_e\omega c \gtrsim 1$ then $W'\gg m_ec^2$. In this limit, the energy-momentum in $L'$ is $p'_{\mu}\simeq (W',W'/c,0,0)$ and thus $p_{\mu}\simeq (\gamma_f W',W'/c,\gamma_f W'/c,0)$ in $L$. The final energy and emission angle $\alpha_{e}$ are given by
 \begin{eqnarray}
{\cal E}_f \simeq  \frac{eE_{\mbox{\tiny SP}}\gamma_f^2}{k}  \simeq m_ec^2 a_{\mbox{\tiny SP}}\alpha \; , \qquad \tan\phi_e=\frac{p_x}{p_y}\simeq \gamma_f^{-1} \; .
\label{eq:theory2}
\end{eqnarray}
Thus, strongly relativistic electrons (${\cal E}_f\gg m_ec^2$) are emitted at small angles $\phi_e$, i.e. close to the target surface. The acceleration length $\ell_{\mbox{\tiny a}}\equiv{\cal E}_f/eE_{\mbox{\tiny SP}}\simeq \lambda\alpha/2\pi$, showing that electrons may reach the highest energy over a few microns. 

The experiment was carried out at the CEA Saclay Laser Interaction Center (SLIC) facility with the UHI100 laser system (see \cite{doi:10.1117/12.2178816} for a preliminary presentation of the experimental results). UHI100 provides a 25 fs laser pulse with a peak power of 100 TW and a wavelength $\lambda\sim 0.8~\mu\textrm{m}$. 
A double plasma mirror \cite{Levy:07} yielded a very high pulse contrast ($\sim 10^{12}$). The pulse was focused on target with an off-axis f/3.75 parabola in a focal spot of $\simeq 4~\mu\mbox{m}$ FWHM containing $\sim 60\%$ of the total energy  in the $1/e^2$ spot diameter, which lead to an average intensity of $\sim 5\times 10^{19}~\textrm{W}/\textrm{cm}^2$. Focal spot optimisation was performed with an adaptive optical system. $P$-polarization was used throughout the experiment. 
\begin{figure}[htb]
\centering
\includegraphics[width=1.0\columnwidth]{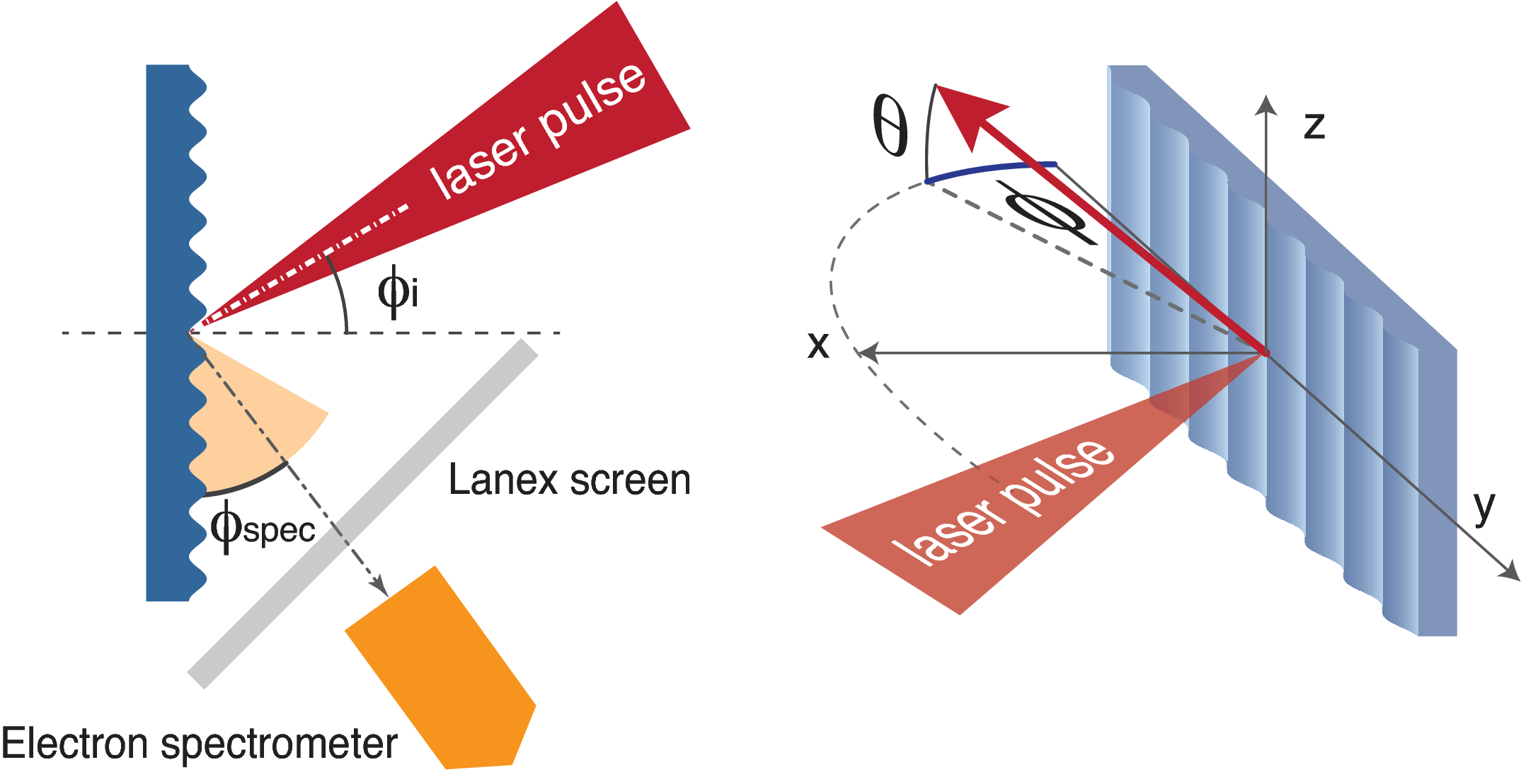}
\caption{Schematic view of the experimental setup. The 2D top view (left) shows the position of the diagnostics.
The 3D sketch (right) shows the adopted convention for the angles $\phi$ and $\theta$.}
\label{fig:expsetup}
\end{figure}
The schematic view of the experimental setup is shown in Fig.\ref{fig:expsetup}.
A compact CMOS based spectrometer, specifically designed for this experiment, was mounted on a motorised tray able to change the angle $\phi_{\mbox{\tiny spec}}$ within the range $0^\circ-60^\circ$ while remaining aligned to the interaction center. The entrance lead pinhole had a diameter of 500~$\mu$m and was placed at 8~cm from the interaction point. A pair of permanent magnets dispersed the electrons on the large ($49.2 \times 76.8 ~ \textrm{mm}^2$ ) triggered 12bit CMOS with 48~$\mu$m pixel size. The energy detection range was $\sim 2-30$~MeV. A scintillating Lanex screen ($16\times5~\mbox{cm}^2$) was used to collect the spatial distribution of the electrons in the full angular range $\phi=0^{\circ}-90^{\circ}$. The screen was placed with an angle of $45^{\circ}$ with respect to the target and its center was at 8~cm from the interaction point. The green light emitted by the Lanex  was selected using a 546 nm band pass filter and recorded by a 12 bits CCD. When in use, the Lanex screen excluded the electron spectrometer. In addition to the electron diagnostics, a Thomson parabola was used to detect protons emitted from the target rear along the normal to the surface, as in a previous experiment \cite{PhysRevLett.111.185001}. The cut-off energy of protons was used as a reference to optimize the focal position of the target.

The grating targets were produced at Czech Technical University, Prague by heat embossing of Mylar{\texttrademark} foils using a metallic master. The target material was chosen considering its high damage threshold for prepulses. In the following we show results obtained with targets having a resonant angle of $\simeq 30^{\circ}$, i.e. $\lambda_g=2\lambda$ having assumed $\omega_p\gg\omega$. The  average thickness was 10~$\mu$m and the peak-to-valley depth of the grooves 0.25~$\mu$m. Flat foils with the same average thickness were used for comparison. 
In a limited number of shots also gratings with a resonance angle of $15^{\circ}$ ($\lambda_g=1.35 \lambda$) and $45^{\circ}$ ($\lambda_g=3.41 \lambda$) were used, obtaining similar results. 
\begin{figure}[htb]
\centering
\includegraphics[width=0.8\columnwidth]{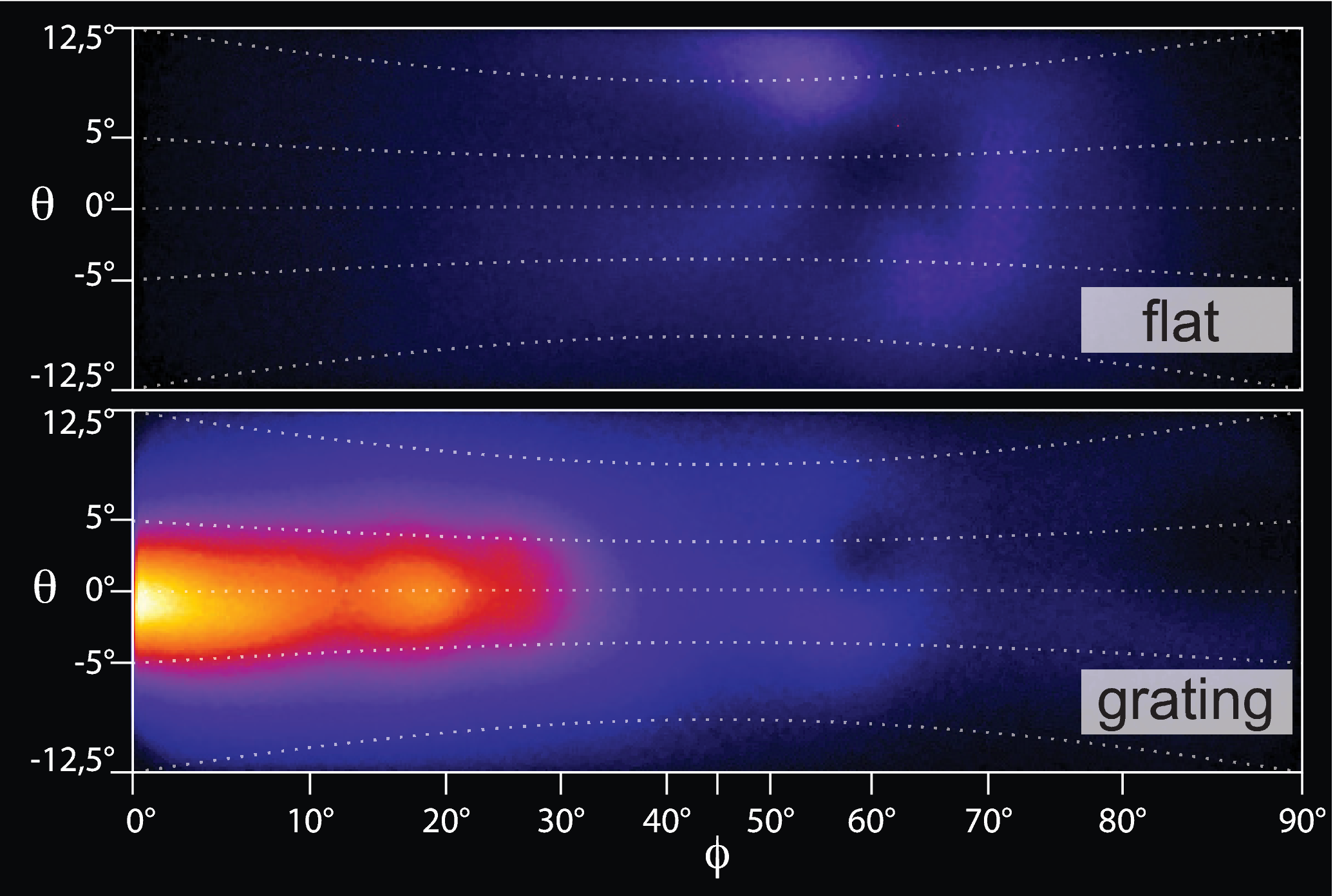}
\caption{Images on the Lanex screen for simple flat target (top) and for grating target (bottom), both irradiated at $\phi_i=30^{\circ}$ incidence. A 3mm Al foil was placed in front of the screen to filter out electrons with energy $E < 1.7~$MeV. The parabolic dashed lines give the local $\theta$ angle corresponding to the position on the screen.}
\label{fig:lanex}
\end{figure}

The electron emission from the front face of the target changes dramatically between gratings irradiated at angles near resonance and flat foils. 
Fig. \ref{fig:lanex} shows the spatial distribution of the electrons, as collected by the Lanex screen. The emission from the flat foil is rather diffused, with a ``hole'' in correspondence of the specular reflection of the pulse, as if electrons were swept away by intense light. The signal is maximum in an annular region around the hole. In contrast, for a grating at resonance the emission is strongly localised on the plane of incidence ($\theta \sim 0^{\circ}$). The maximum intensity is close to the target tangent and is  $\sim 10$ times larger than what observed for flat targets at the same angle of incidence. Two minima (holes in the image) are observed in the directions of specular reflection and  first-order diffraction of the laser pulse (evidence of grating diffraction of the high-intensity pulse, which confirms the survival of the grating during the interaction, was also found in previous measurements \cite{PhysRevLett.111.185001}).
Local bending of the target or non-exact perpendicularity  of the grating grooves to the plane of incidence may result in shot-to-shot fluctuations of the direction of maximum emission. Depending on the individual foil, the average angular shift in $\theta$ was in the $1^{\circ} - 5^{\circ}$ range.
An optimization of target design and alignment is foreseen to eliminate the fluctuations.

\begin{figure}[htb]
\centering
\includegraphics[width=0.95\columnwidth]{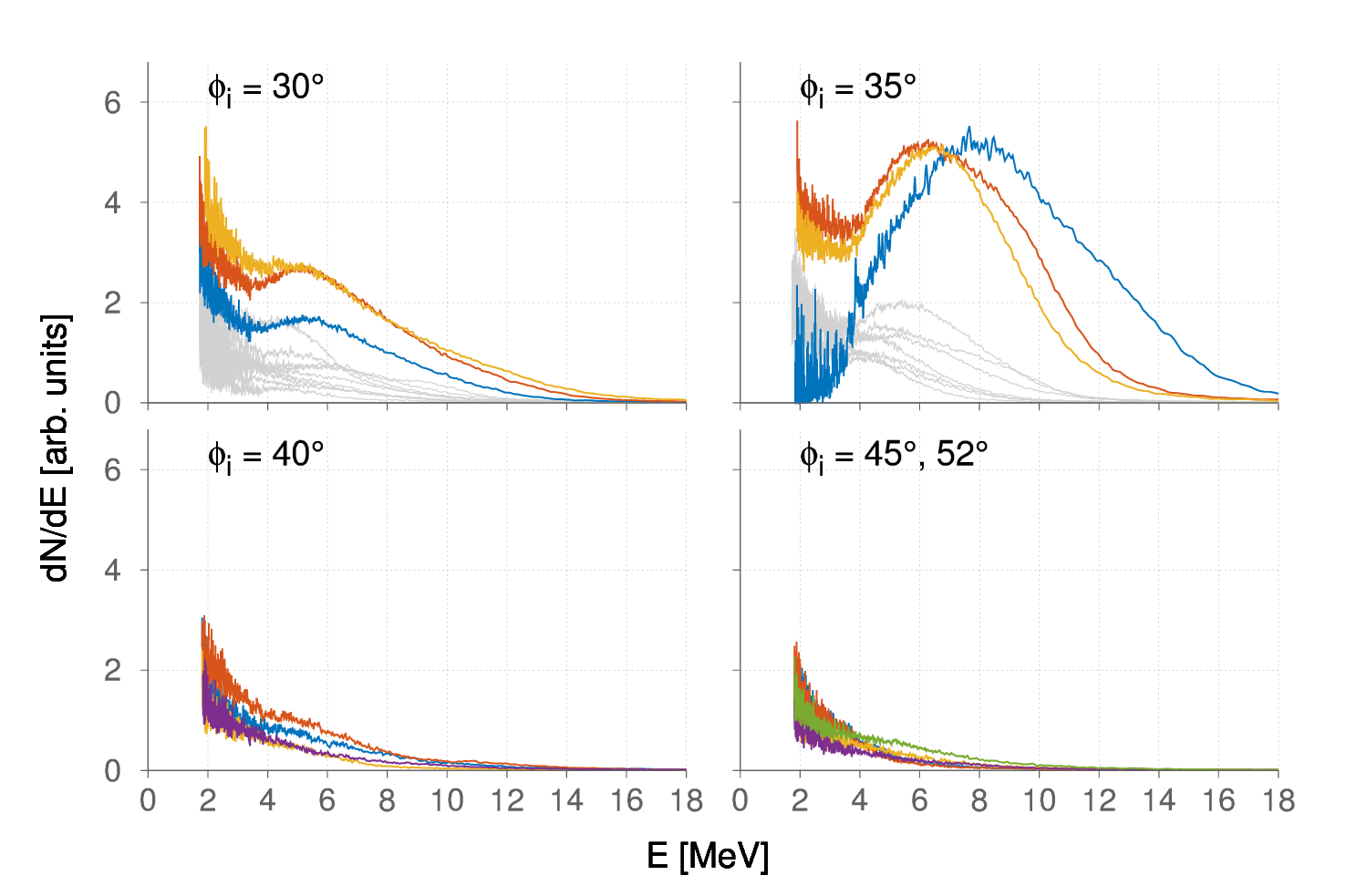}
\caption{Electron spectra collected with the detector at 2$^{\circ}$ from tangent direction, for several pulse incidence angles (from 30$^{\circ}$ to 52$^{\circ}$). In the upper panels, all the collected shots are shown and a few of them are highlighted in color, while the others are plotted in light grey to show the shot to shot variations.}
\label{fig:multiSpec}
\end{figure}
The energy spectra were obtained  placing the spectrometer  at $\phi_{\mbox{\tiny spec}} = 2^{\circ}$. The angle of incidence of the laser $\phi_i$ was varied from $20^{\circ}$ to $52^{\circ}$. Fig. \ref{fig:multiSpec} shows spectra obtained for  $\phi_i\geq30^{\circ}$ as for smaller angles no clear signal above the noise level was collected. The aforementioned fluctuations of the direction of the electron beam lead to a shot-to-shot variability of the intensity of the signal. Nevertheless, the most intense signals are detected only close to the resonance angle ($\sim 30^{\circ}$). Moreover, spectra collected at  $30^{\circ}$ and  $35^{\circ}$ are characterized by higher maximum energies and a peculiar distribution with a dip at lower energies (3-4~MeV) and a broad peak at 5-8~MeV. On the higher energy side electrons with energy up to $\sim 20~\mbox{MeV}$ are detected. Whichever the angle of incidence, we never observed an electron spectrum above the noise level from flat targets at $\phi_{\mbox{\tiny spec}} \sim 2^{\circ}$.
We also analysed the electron spectra obtained irradiating the grating at $\phi_i=30^{\circ}$  and changing the position of the spectrometer in the $\phi_{\mbox{\tiny spec}} = 1^{\circ}-35^{\circ}$ range.
Despite the shot-to-shot fluctuation, the shape of the spectra remained similar in all positions of the spectrometer $\phi_{\mbox{\tiny spec}} \lesssim 20^{\circ}$,with a peak at 5-8 MeV. On the other hand, the intensity of the signal monotonically decreased with respect to $\phi_{\mbox{\tiny spec}}$ and the signal was visible up to  $\phi_{\mbox{\tiny spec}} \simeq 30^{\circ}$, in agreement with the signal collected on the Lanex screen.

Massively parallel 3D simulations were performed considering flat targets irradiated at $\phi_i=30^{\circ}$ and gratings irradiated at   $\phi_i=30^{\circ}, 35^{\circ}, 40^{\circ}$.   For reasons of computational feasibility, the thickness $\ell_t$ of the simulated target was $\ell_t=1\lambda$ and the electron density was $n_e=50n_c$ (where $n_c=\pi m_ec^2/e^2\lambda^2$ is the cut-off density),
while the other parameters corresponded to the experimental ones.
The simulations were performed on 16384 cores of the FERMI supercomputer using the open-source particle-in-cell (PIC) code ``PICCANTE'' \cite{piccante,2015arXiv150302464S}. The numerical box size was $80\lambda \times 80 \lambda \times 60 \lambda$, wide enough for the boundaries not to affect the results. A spatial resolution of 70, 51, and 34 points per $\lambda$ in each direction and 50 particles per cell were used. 
\begin{figure}[htb]
\centering
\includegraphics[width=0.9\columnwidth]{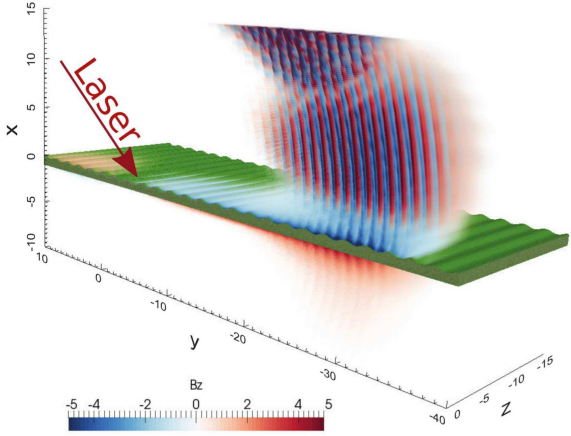}
\caption{3D particle-in-cell simulation snapshot of laser-grating interaction, showing the magnetic field component $B_z$ and an isosurface highlighting the target at time $t=45\lambda/c$ after the beginning of the interaction. Only the $z<0$ region of simulation box is shown in order to highlight the field distribution. In the upper part of the picture, the distribution of $B_z$ is related to the diffracted pulse, while that in the lower part corresponds to a surface plasmon propagating in the $-\hat{y}$ direction.
}
\label{fig:3DPIC}
\end{figure} 

Fig. \ref{fig:3DPIC} shows a snapshot of a 3D simulation of a grating target irradiated at its resonance angle (the interaction geometry is shown in Fig.\ref{fig:expsetup}). The magnetic field component $B_z$ is represented together with the isosurface corresponding to the electron density of the grating target. A surface wave propagating along the $-\hat{y}$ direction is excited and electrons trapped in the SP are accelerated along the target.
\begin{figure}[htb]
\centering
\includegraphics[width=1.0\columnwidth]{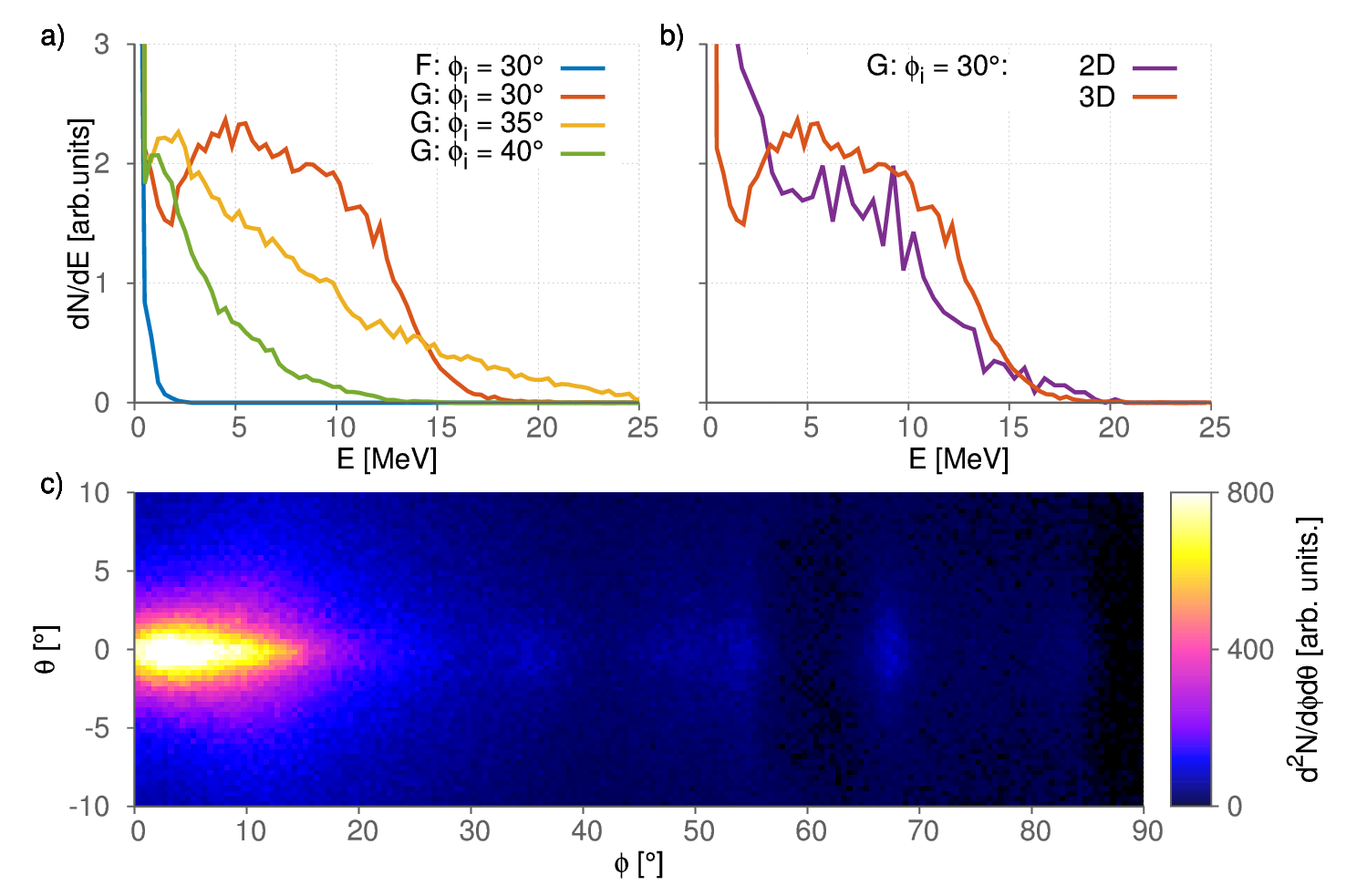}
\caption{Analysis of 3D simulations at time $t=45\lambda/c$ after the beginning of the interaction. a) Simulated spectrum $dN/dE$ at $\phi_{spec}=2^{\circ}$ for flat (F) targets and for gratings (G) at different values of the incidence angle $\phi_i$.  Particles with $x <- 0.125$ (the average surface position of the target),  $\vert\phi - 2\vert > 0.5^{\circ}$ or $\vert\theta\vert > 0.5^{\circ}$ are filtered out.
b) Comparison of spectra between 2D and 3D simulation for a grating target irradiated at $\phi_i=30^{\circ}$.
c) Electron angular distribution $d^2N/d\phi d\theta$ for grating irradiated at $30^{\circ}$. Particles with $x < - 0.125$ or kinetic energy $E_k<1$ MeV are filtered out. 
}
\label{fig:3Dspectra1}
\end{figure} 
Fig. \ref{fig:3Dspectra1}~a) shows the simulated electron energy spectra $dN/dE$ at $\phi_{spec}=2^{\circ}$ for the flat target irradiated at $\phi_i=30^{\circ}$ and grating targets irradiated at $\phi_i=30^{\circ},35^{\circ},40^{\circ}$. The signal for the flat target case is very weak compared to grating spectra and it exhibits an energy cut-off which is $\sim 10\times$ lower. The energy spectrum for the grating irradiated at $\phi_i=30^{\circ}$ shows the peculiar spectral shape observed for $\phi_i=30^{\circ},35^{\circ}$ in the experimental results (see Fig. \ref{fig:multiSpec}), while for higher angles of incidence the low energy dip is not observed as for $\phi_i \geq 40^{\circ}$ in the experiment. 
Fig. \ref{fig:3Dspectra1}~b) compares the spectra obtained in 2D and 3D simulations. Only the 3D simulation fully reproduces details of the spectrum such as the broad peak with low energy dip.
Fig. \ref{fig:3Dspectra1}~c) shows the simulated angular distribution on the screen, which also reproduces fairly well the experimental data (Fig.\ref{fig:lanex}), including the hole in correspondence of the angle of specular reflection.

The 3D simulation also shows a correlation between electron energy and emission angle. Electrons at energies lower than the peak value are emitted at some angle with respect to SP propagation direction, so that integrating over the whole range of $\theta$ the spectrum resembles the one observed in the 2D case. This is consistent with interpreting the fluctuations in the energy spectra (Fig.3) as related to those in the electron beam direction.

In the simulation, $a_{\mbox{\tiny SP}}\simeq 1$ showing that the SP is relativistic. Inserting such value and $\alpha=50$ in Eq.(\ref{eq:theory2}) we obtain ${\cal E}_f \simeq 25~\mbox{MeV}$, $\phi_e \simeq 8^{\circ}$ and $\ell_{\mbox{\tiny a}}\simeq 8\lambda$, in fair agreement with the observations. The linear scaling of ${\cal E}_f$ with $a_0$ is also apparent in parametric 2D simulations. Details of the spectrum will be dependent on the distribution of injection velocities. In the supplemental material, movies from 2D simulations showing the acceleration of electrons moving along the surface are reported.
In a very recent work which explores electron accerelation
regimes in surface plasma wave \cite{riconda_popinpress}, possible self-injection  and phase-locking of electrons at relativistic intensity in the surface plasma wave is shown with a test particle approach.

Our results may have application as an intense ultra-short electron source in the multi-MeV range, with characteristics not easily attainable with other techniques. Electrons with these parameters are potentially of interest for photoneutron generation \cite{PhysRevLett.86.2317,PhysRevLett.113.184801,JapanNeutrons} (aiming at extremely high fluxes) or ultra-fast electron diffraction \cite{0034-4885-74-9-096101,:/content/aip/journal/apl/95/11/10.1063/1.3226674,:/content/aip/journal/apl/89/18/10.1063/1.2372697} (for imaging of ultrafast processes with electron diffraction). Besides these possible applications our results confirm the possibility to extend the study and applications of plasmonics into the high field, relativistic regime.

\begin{acknowledgments}
The research leading to these results has received funding from LASERLAB-EUROPE (grant agreement no. 284464, EU's Seventh Framework Programme.
Support from 
``Investissement d'Avenir'' LabEx PALM (Grant ANR-10-LABX-0039), Triangle de la physique /contract nbr. 2014-0601T ENTIER) and ``Institut Lasers et Plasmas'' is also acknowledged.
Partial support of Czech
Science Foundation project No. 15-02964S is gratefully acknowledged. 
We acknowledge ISCRA and LISA access schemes to the BlueGene/Q machine FERMI, based in Italy at CINECA, via the projects ``FOAM2'' (ISCRA) and ``LAPLAST'' (LISA), and PRACE for the development of the code within the preparatory access project ``PICCANTE''.
The authors would like to thank the team of Saclay Laser Interaction Center for their support, and Ondrej Klimo, Fabien Qu\'er\'e, Gianluca Sarri and Caterina Riconda for useful help and discussions.
\end{acknowledgments}



\date{\today}
\end{document}